\documentclass[epj,nopacs]{svjour}

\usepackage{parskip,amsmath,graphicx,tabulary,floatrow,bm, amssymb,textcomp}
\usepackage{graphicx}
\usepackage{color}

\begin{document}

\title{Influence of quantum fluctuations on the superfluid critical velocity of a one-dimensional Bose gas}
\titlerunning{Influence of quantum fluctuations on the superfluid critical velocity}

\author{Chao Feng\inst{1}\and Matthew J. Davis\inst{2}}

 \institute{School of Mathematics and Physics, The University of Queensland, Brisbane, Queensland 4072, Australia \and 
 ARC Centre of Excellence in Future Low-Energy Electronics Technologies, School of Mathematics and Physics, The University of Queensland, Brisbane, Queensland 4072, Australia}

\mail{mdavis@physics.uq.edu.au}
\abstract{The mean-field Gross-Pitaevskii equation with repulsive interactions exhibits frictionless flow when stirred by an obstacle below  a critical velocity.  Here we go beyond the mean-field approximation to examine the influence of quantum fluctuations on this threshold behaviour in a one-dimensional Bose gas in a ring.  Using the truncated Wigner approximation, we perform simulations of ensembles of trajectories where the Bose gas is stirred with a repulsive obstacle below the mean-field critical velocity.  We observe the probabilistic formation of grey solitons which subsequently decay, leading to an increase in the momentum of the fluid.  The formation of the first soliton leads to a soliton cascade, such that the fluid rapidly accelerates to minimise the speed difference with the obstacle.  We measure the initial rate of momentum transfer, and relate it to  macroscopic tunnelling between quantised flow states in the ring.
}

\maketitle

\section{Introduction}

The experimental control and manipulation of dilute gas  Bose-Einstein condensates (BECs) has advanced rapidly over the last decade. A number of potential applications of BECs  such as atomtronic circuits~\cite{Jendrzejewski,Eckel} and superconducting quantum interference devices (SQUIDs)~\cite{Levy,Wright,Wright2,Murray} rely on their superfluid properties.  Going beyond proof of concept experiments, however, will require a precise understanding of how the superfluid properties of a BEC are affected by external perturbations and other influences such as quantum fluctuations. Examples include understanding when the weak link in an atomic SQUID will lead to the decay of the superfluid current~\cite{Eckel,Jendrzejewski,Ramanathan,Kumar}; how superfluidity is affected by the manipulation of  barriers and confining potentials~\cite{Mathey,Feng,Watanabe,Hakim,Engels,Pavloff,Piazza,Leboeuf,Moritz,Singh1,Singh2,Polo};  and when the imperfections of an atomic waveguide will impede  superfluid flow~\cite{Moulder}. There is also interest in the controlled generation of quantum phase slips --- excitations in dissipationless superfluids --- to create topologically protected qubits and standardised quantum currents~\cite{Tanzi}, and in the modulation and control of superfluid parameters~\cite{Higbie,Onofrio_2012}.

One of the defining characteristics of a superfluid is the critical velocity $v_c$: the speed above which any perturbation moving within the superfluid will experience a drag force and lead to excitations.  When the critical velocity is exceeded when stirring a superfluid with a repulsive obstacle, dark solitons (in 1D) and vortices (in 2D and 3D) are the dominant form of excitations, generally also accompanied by dispersive waves~\cite{Hakim,Pavloff,Jackson_Simulate_Raman,Jackson_Vortex,Crescimanno,Feng,Singh1,Singh2}.

Within mean-field theory, the critical velocity can be estimated using the Landau criterion.  For weakly-interacting dilute gas BECs with repulsive interactions the critical velocity is predicted to be equal to the local speed of sound~\cite{Watanabe,Feng}.  Experiments have demonstrated the existence of a critical velocity in BECs~\cite{Onofrio,Raman,Engels,Neely,Moritz} and a large body of theoretical work has examined this phenomenon based on the mean-field Gross-Pitaevskii equation (GPE) --- see, for example, Refs.~\cite{Hakim,Astrakharchik,Watanabe,Danshita,Jackson_Simulate_Raman,Jackson_Vortex,Frisch,Leboeuf,Leszczyszyn,Pavloff,Crescimanno,Feng,Syafwan,Huepe}. These have found that while the critical velocity observed in experiments is often smaller than the bulk speed of sound in the BEC, this can sometimes be reconciled with the Landau criterion if one carefully takes into account the local conditions around the obstacle~\cite{Feng,Watanabe}.  This involves estimating a local speed of sound and a local flow velocity which depend upon the local density, the depth and shape of the perturbing potential, and the speed of the obstacle. Other authors have made use of classical field simulations for BECs~\cite{Blakie}, and found that the effects of finite temperature, inhomogeneous density, and circular stirring all contribute to the reduction of critical velocity from the speed of sound~\cite{Mathey,Singh1,Singh2}.  Katsimiga \emph{et al.} have studied the creation of solitons due to an oscillating obstacle in a 1D harmonically trapped BEC using beyond mean-field methods~\cite{Katsimiga2018}.

The existence of a critical velocity for superfluid flow leads to the concept of persistent currents in multiply-connected geometries such as a ring trap.  A superflow moving at a subcritical speed can persist indefinitely, as no excitations can form to reduce the momentum of the fluid~\cite{Leboeuf,Pavloff,Hakim}.  The superflow is in fact a metastable state --- it is not the thermodynamic ground state of the system, but is classically prevented from reaching the ground state due to the existence of an energy barrier~\cite{Coleman,Kagan,Buchler,Kumar}.  

Beyond mean-field theory, at zero temperature quantum tunnelling can lead to the decay of the superflow towards the thermodynamic ground state~\cite{Coleman,Leggett}.  The energy states of this system are a series of progressively lower energy minima separated by barriers. The transition between two consecutive minima is pedagogically similar to the textbook calculation of the transition between the minima of an asymmetric double well~\cite{Coleman,Leggett,Sethna}. The minima of the wells  correspond to quantised flow states, labelled by integer quantum numbers $q$ and $q'$, separated by energy $\Delta{E}=E_{q'}-E_q$.  At finite temperatures the decay of the superfluid can also be caused by thermal activation~\cite{Snizhko}.


A number of theoretical studies have calculated the rate of decay of a superfluid disturbed by a moving obstacle~\cite{Kagan,Cherny,Cherny_Review,Buchler}. A common approach has been to calculate the transition rate from an initial metastable flow state $q$ to another flow state $q'$. The most widely cited approaches are the instanton method~\cite{Buchler,Sethna,Coleman,Leggett,Danshita} and effective Hamiltonian methods~\cite{Kagan,Kolovsky,Polkovnikov}, although results have also been obtained by applying the Bethe Ansatz approach to the Lieb-Liniger model~\cite{Cherny}.

In the instanton method, the two flow states are approximated as the minima of an asymmetric double-well system, and the tunnelling rate from one well to the other can be calculated using well established path integral methods~\cite{Buchler}. This technique has been used to calculate the transition probability from higher to lower superfluid flow states for a subcritical toroidal flow in the presence of a  delta-like obstacle~\cite{Buchler} and for an array of such obstacles~\cite{Danshita}. A characteristic finding of these analytical studies is the prediction of a decay rate that is a power-law function of the velocity. 

These previous theoretical works in one-dimension focused on the initial rate of decay of the superflow, with limited discussion of the microscopic dynamics of the density, or the long time behaviour of the system.  In this paper we address this gap in the literature.  We consider a homogeneous one-dimensional Bose gas at zero temperature with periodic boundary conditions as could be realised experimentally in a ring trap with tight transverse confinement. A number of experiments in similar configurations have provided evidence for dissipative effects that are not readily described by mean-field simulations~\cite{Kumar,Eckel,Jendrzejewski,Ramanathan}. Some have suggested that probabilistic vortex formation could play a role~\cite{Kumar,Jendrzejewski}.

Here we use the truncated Wigner approximation to examine the influence of quantum fluctuations on the dynamics of a stationary Bose gas when it is stirred by a moving obstacle at a speed $v$. We initially focus on the  situation when the obstacle moves at speeds  below the mean-field critical velocity $v_c$ as determined by simulation of the Gross-Pitaevskii equation for the same system parameters.  In particular, we examine how quantum fluctuations result in the probabilistic occurrence of phase slips, and the formation of gray solitons  in the TWA simulation trajectories for obstacle speeds \emph{below} the mean-field critical velocity.  We quantify the excitations of the system by computing the increase in momentum over a fixed time interval,
and find that there is no sharp transition at the mean-field critical velocity.

For the long time dynamics we find that the initial creation of a soliton leads to a rapid cascade of further solitons so that the eventual difference in speed between the stirrer and the fluid is minimised. We calculate the rate of increase of the fluid momentum as a function of the obstacle velocity and interaction strength, and compare these to earlier analytical approximations of superfluid decay in one-dimension~\cite{Kagan,Buchler,Cherny,Cherny_Review,Danshita}.

\begin{figure*}
\includegraphics[width=0.92\textwidth]{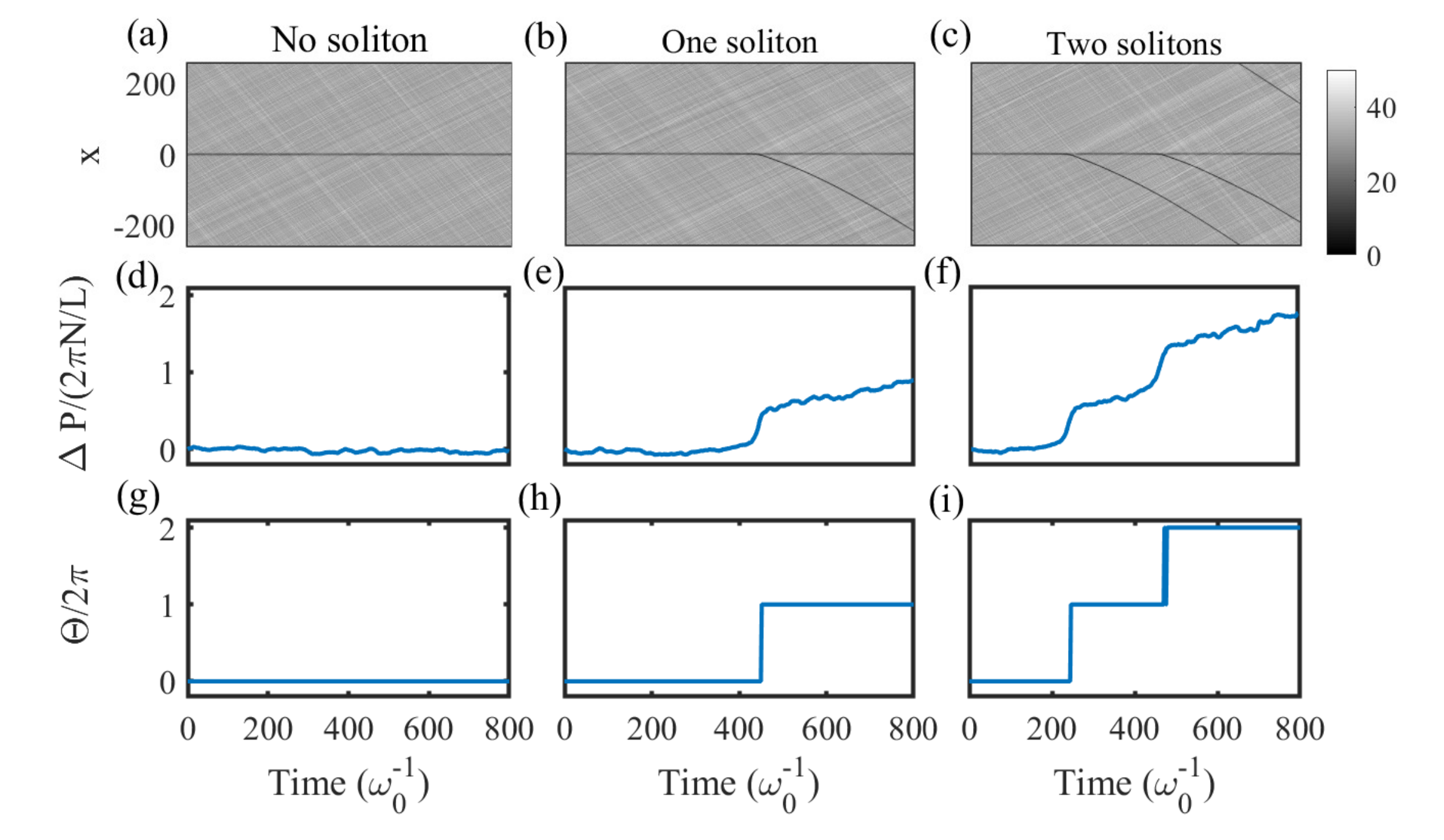}
\caption{Exemplar trajectories showing probabilistic soliton formation in TWA simulations below the critical velocity. (a--c) shows  $|\psi(x,t)|^2$ as a function of time for trajectories in which zero, one, and two solitons form, respectively. (d--f) The change in momentum, and (g--i) the system phase winding as a function of time  for the same trajectories.  Simulation parameters are $g=0.025$, $N=16384$, $L=512$, $512$ grid points, $V_0=0.3\mu$ and $v=0.98646v_c$.
\label{fig:single_trajectories}}
\end{figure*}

\section{Model}

The system of interest is a stationary homogeneous one-dimensional Bose gas with periodic boundary conditions (i.e.~a ring-like geometry).  It is subjected to stirring by a penetrable obstacle moving with a velocity $v$ that is below the mean-field critical velocity.  
We perform a Galilean transformation to perform simulations in the frame of reference in which the obstacle is stationary at $x=0$.

We work in dimensionless units $x' = x/x_0$, where $x_0 = \sqrt{\hbar/m\omega_0}$, $t' = \omega_0 t$,  and $\omega_0$ is the  trapping frequency of an assumed harmonic potential in the transverse dimensions that allows the one-dimensional reduction.  The dimensionless interaction parameter is $g'=4\pi a_s/x_0$ where $a_s$ is the s-wave scattering length. From this point forward, dashes on these dimensionless quantities in our notation are dropped for clarity. 

The time-dependent GPE in the frame moving with the obstacle is
\begin{equation}
i\frac{\partial \psi(x)}{\partial t} =  \left( -\frac{1}{2}\frac{\partial^2 }{\partial x^2} - iv\frac{\partial }{\partial x} + V(x)   + g|\psi(x)|^2\right) \psi(x),
\label{eq:moving_dimensionless_GPE_1D}
\end{equation}
where $\psi(x)$ is the classical Bose field normalised to the number of particles $N = \int dx |\psi(x)|^2$.  The potential has the form
\begin{equation}
V(x) = V_0 \exp( - 2x^2/\sigma^2).
\end{equation}
We consider the situation where the  spatial extent of the obstacle $\sigma = 8\xi$, where the healing length $\xi=(2\mu)^{-1/2}$, and $\mu = g n$ is the chemical potential with $n$ the background density, such that the system is in the hydrodynamic regime.  The potential height  is chosen to be $V_0 = 0.3 \mu$, a  moderate fraction of the system chemical potential such that the obstacle is penetrable.  Simulations with smaller potential heights (i.e. $\ll 0.3\mu$) lead to qualitatively similar results, but with a slower soliton formation rate.  This has the practical effect that significantly longer simulation times are required in order to obtain the same level of statistical significance in the results.  Conversely, larger potential heights (i.e. $\gg 0.3 \mu$) can adversely affect the flow conditions in the region of the obstacle, and can lead to effects beyond the scope of this study, such as shockwave formation, or the pinning of the condensate~\cite{Syafwan}. We chose $V_0 = 0.3 \mu$ as a convenient middle ground.

To explore the beyond-mean-field effects due to quantum fluctuations, we numerically simulate this system using the TWA~\cite{Blakie}.  Briefly, in this approach the time evolution of the full quantum field is approximated by  a stochastic ensemble of initial classical fields $\psi(x)$ evolved according to the Gross-Pitaevskii equation~(\ref{eq:moving_dimensionless_GPE_1D}). The method is valid in the weakly-interacting regime where the number of simulation modes $M$ is significantly less than the number of particles $N$, and for  simulation times such that the neglect of the higher-order terms in the equation of motion for the Wigner function is reasonable.

In this paper we only consider stationary zero-temperature initial states, which are constructed according to
\begin{equation}
    \psi(x) = \psi_0(x) + \sum_i^M \left[ \eta_i u_i(x) + \eta_i^* v^*_i(x)\right]\
\end{equation}
where $\psi_0(x)$ is the time-independent ground state of the Gross-Pitaevskii equation~(\ref{eq:moving_dimensionless_GPE_1D}) and $u_i(x)$ and $v_i(x)$ are the solutions for the Bogoliubov excitations of the system.  $\eta_i$ are complex Gaussian random numbers with $\langle|\eta_i|^2\rangle = 1/2$, such that each Bogoliubov mode has an ensemble average occupation of half a particle.  Further details of this methodology are described in Refs.~\cite{Blakie,Martin}. 

Stationary solutions to Eq.~(\ref{eq:moving_dimensionless_GPE_1D}) are only possible for obstacle speeds lower than the critical velocity $v<v_c$.  Other typical parameters for the simulations are interaction strength $g=0.1$, number of atoms $N=2048$,  system length $L=128$ and number of modes $M=128$. 

Ensemble averages are based on between 96 to 3000 trajectories, as needed to attain a suitably small statistical uncertainty~\footnote{Where comparisons are made across different simulation parameters, the same number of trajectories were used to ensure statistical consistency.}. An analysis of statistical uncertainty was performed for our simulations and the 95\%  confidence interval for the uncertainty in the mean is indicated on relevant figures.

For the purposes of this paper it is important that the mean field critical velocity $v_c$ is determined to high numerical accuracy. Previous studies~\cite{Watanabe,Hakim,Pavloff,Leboeuf,Feng}  have shown that the critical velocity depends sensitively on the local density and is modified by local flow conditions. It is also affected by the width and height of the obstacle, and its velocity relative to the local flow speed~\cite{Watanabe,Pavloff}. Watanabe \emph{et al.}~\ {Watanabe} showed for a homogeneous system in the hydrodynamic regime, the critical velocity can be determined by solving the Bernoulli equation at the threshold indicated by the Landau criterion. In our units, the results in Ref.~\cite{Watanabe} can be cast as
\begin{equation}
\left( \frac{v}{\sqrt{2}c} \right)^2 - \left(\frac{3}{2}\right)^{\frac{2}{3}} \left( \frac{v}{\sqrt{2}c} \right)^{\frac{2}{3}} + 1 - \frac{V_0}{\mu} = 0,
\label{eq:paper_watanabe_velocity}
\end{equation}
where $V_0$ is the maximum amplitude of the impurity potential and $\mu = c^2$ is the chemical potential. Physically, the smaller of two solutions to Eq.~(\ref{eq:paper_watanabe_velocity}), $v_-$, corresponds to the critical velocity, while the larger solution $v_+$ is the speed above which no energy is transferred to the system. In general, $v_c\equiv{v_-}$ is lower than the speed of sound calculated as $c=\sqrt{gn}$ --- in our case a typical value is $v_c\approx 0.44c$.

The estimate Eq.~(\ref{eq:paper_watanabe_velocity}) is exact in the hydrodynamic limit of $\sigma/\xi\rightarrow\infty$. In our case $\sigma/\xi=8$ and Eq.~(\ref{eq:paper_watanabe_velocity}) is only accurate to one or two significant figures. We therefore numerically determine $v_c$ to 8 or 9 significant figures using an iterative algorithm to test whether stationary solutions are possible, and check by propagating in real time using  Eq.~(\ref{eq:moving_dimensionless_GPE_1D}). 

\section{Results}

\subsection{Behaviour of individual trajectories}

It is instructive to first examine individual trajectories of an ensemble of simulations in the truncated Wigner approximation. Consider a system that is stirred at a speed of $v=0.98646v_c$, where the critical velocity was numerically determined to be $v_c = 0.39442$, and other parameters as specified in the caption of Fig.~\ref{fig:single_trajectories}.

Grey soliton formation is a well-known phenomenon in GPE simulations above $v_c$~\cite{Hakim,Watanabe,Pavloff,Feng}, but here we find solitons form in some trajectories \emph{below} $v_c$.
While individual trajectories do not have a well-defined physical interpretation, it has previously been argued that they can loosely be interpreted as single realisations of experiments~\cite{Norrie,Blakie,Weiler,LewisSwan}.  In the simulations each trajectory has a different realisation of noise in the initial state, and we find that this leads to macroscopically different outcomes after some simulation time.

This behaviour is illustrated in three exemplar trajectories in Figs.~\ref{fig:single_trajectories}(a--c), which show the evolution of $|\psi(x,t)|^2$, roughly corresponding to density. The stationary dark depression at $x=0$ in each of Figs.~\ref{fig:single_trajectories}(a--c) is caused by the stirrer, which is stationary in the simulation frame of reference. The key distinguishing feature between the different trajectories is the emergence, in some trajectories, of persistent and dark density depressions which are identified as grey solitons.

Figure~\ref{fig:single_trajectories}(a) shows a trajectory in which no solitons form during the simulation time.  Figure~\ref{fig:single_trajectories}(d) shows the corresponding change in total system momentum, which exhibits only small fluctuations about zero. Figures~\ref{fig:single_trajectories}(b) and (c), on the other hand, show the formation of one and two solitons respectively, with corresponding step-like changes in the system momentum (shown in Figs.~\ref{fig:single_trajectories}(e) and (f)). The momentum changes occurs in initial discrete jumps of approximately $\pi{N}/L$ and then rapidly increases to around $2\pi{N}/L$, corresponding to the difference in momentum between quantised superflow states $q$ and $q-1$, where $q$ is an integer.

To highlight the direct connection between soliton formation and change in momentum, we calculated the phase winding of individual trajectories using $\Theta = \int_0^L (d\theta(x)/dx) dx$, where $\theta(x)$ is the unwrapped phase of the classical field $\psi(x)$, as a function of time [see Figs.~\ref{fig:single_trajectories}~(g--i)].  Every grey soliton formation event and corresponding change in system momentum results in a $2\pi$ phase slip.

\subsection{Soliton formation and ensemble momentum}
\begin{figure}
\includegraphics[width=0.90\textwidth]{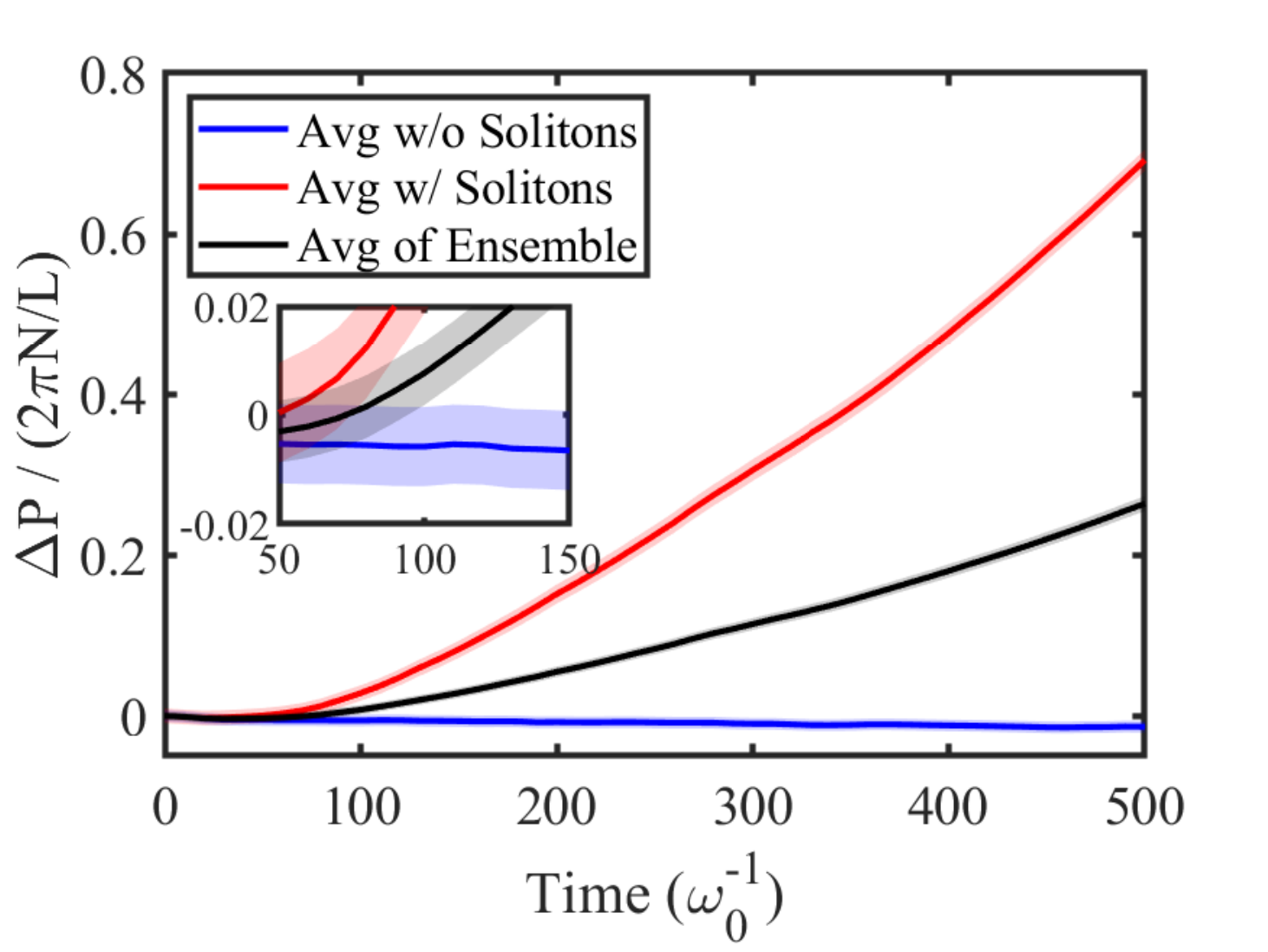}
\caption{Momentum change averaged over three ensembles: 1821 trajectories where no soliton formation was observed (blue); 1179 trajectories where at least one soliton has formed (red), and all 3000 trajectories (black). The shaded regions represent the 95\% confidence interval, which are similar to the line width (see inset). All simulation parameters are the same as in Fig.~\ref{fig:single_trajectories}.
\label{fig:ensemble_momenta}}
\end{figure}

While the soliton formation in individual trajectories occurs stochastically, the expectation value of the momentum
\begin{equation}
    P(t) = \bigg\langle 
   -i \int  \psi^*(x) \frac{d\psi(x)}{dx}  dx \bigg\rangle,
\end{equation}
of the ensemble increases smoothly, and is plotted in Fig.~\ref{fig:ensemble_momenta}. This is similar to results of Refs.~\cite{Mathey,Danshita} for the decay of an initial superfluid flow from the probabilistic nucleation of phase slips due to quantum and thermal fluctuations; see also Ref.~\cite{Polo}. 

To highlight that soliton formation is responsible for the change in  momentum $\Delta{P} = P(t_f) - P(0)$, we divided the trajectories into two ensembles containing (i) those in which  solitons  formed, and (ii) those in which solitons have not formed within the simulation time.  As can be seen in Fig.~\ref{fig:ensemble_momenta}, the average momentum of the non-soliton forming ensemble does not show any appreciable increase~\footnote{We note there is a small change due to the initial equilibration of the truncated Wigner initial state.}. By contrast, the average momentum change for the soliton-forming ensemble is significant.

\begin{figure*}
\includegraphics[width=0.9\textwidth]{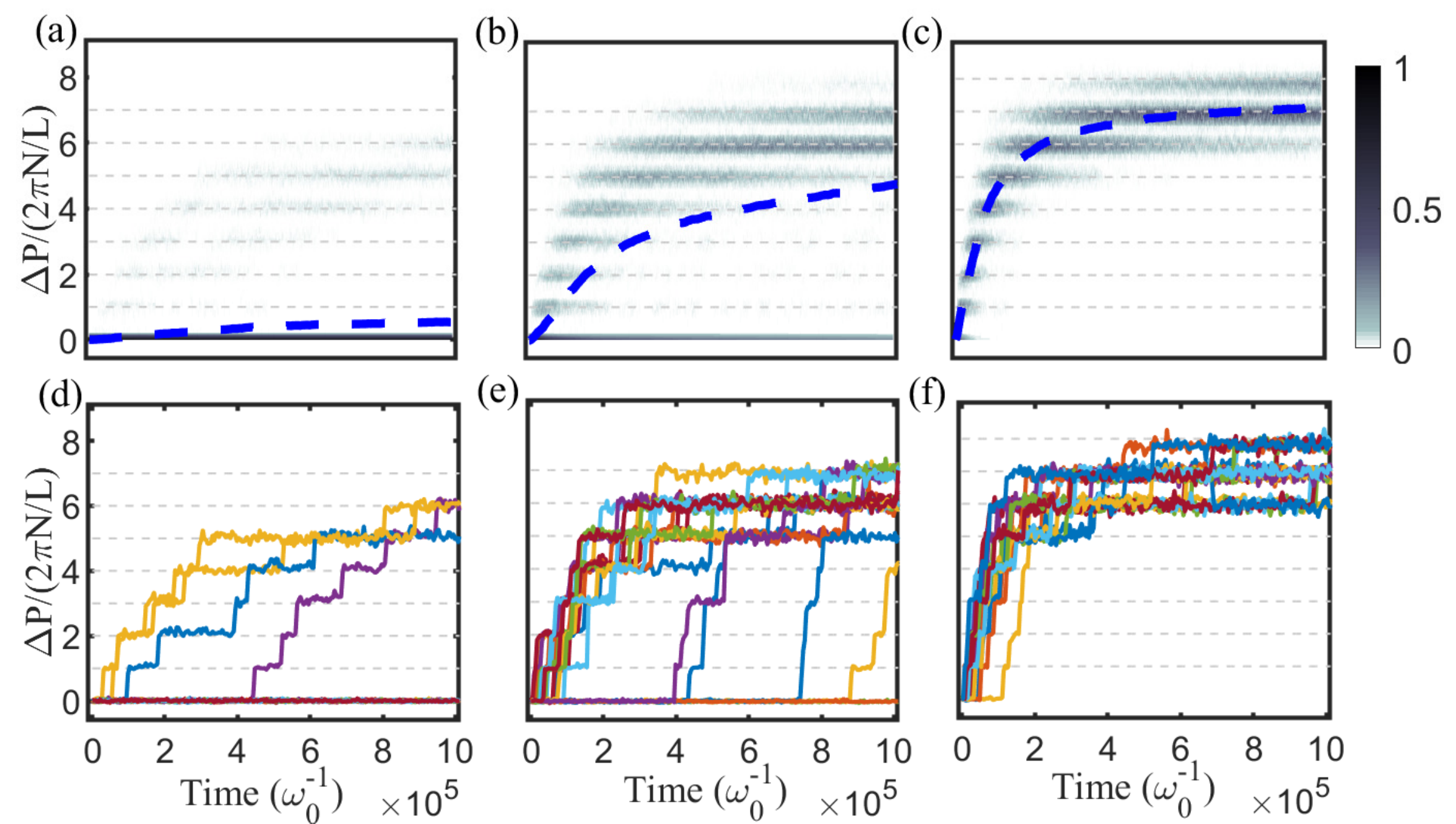}
\caption{(a--c) Density plots showing histograms of the change in momentum as a function of time over the ensemble of trajectories for three different stirring velocities (a) $v/v_c=0.88$, (b) $v/v_c= 0.92$, and (c) $v/v_c=0.97$, sorted into 40 bins over $\Delta{P}/(2\pi{N}/L)\in[0,8]$. The blue dashed line show the ensemble average momentum change for all 248 trajectories. (d--f) Momentum change for a random sample of thirty trajectories from the ensemble for the same set of velocities as in (a--c). In (d) and (e) several trajectories have no change in momentum during the simulation time of $10^6\omega_0^{-1}$. Simulation parameters are $g=0.025$, $N=4096$, $L=128$ using $128$ grid points and $V_0=0.3\mu$. 
\label{fig:trajectories_long}}
\end{figure*}

As the stirring velocity $v$ increases toward $v_c$, the fraction of trajectories exhibiting soliton formation at a given time increases significantly, as does the rate of momentum change. This is consistent with the theoretical understanding that the rate of tunnelling of the fluid between quantised flow states will increase as $v\rightarrow{v_c}$.  We quantitatively address this aspect in Section~\ref{sec:decay_rate}.

Figures~\ref{fig:trajectories_long}(a--c) show histograms of the change in momentum as a function of time over the ensemble of trajectories  for three different stirring velocities $v/v_c=0.88$, $0.92$ and $0.97$. In Figs.~\ref{fig:trajectories_long}(a,b), where $v/v_c=0.88$ and $0.92$ respectively, only a fraction of the trajectories exhibit soliton formation in the simulation time frame, and of those that do, not all have reached the final momentum state of ${\Delta}P_0=Nv$. At long times, a significant fraction remain in the $\Delta{P}=0$ state.

In Fig.~\ref{fig:trajectories_long}(c) where $v/v_c=0.97$, it can be seen that all trajectories exhibit soliton formation, and the total momentum change over the interval is close to $\Delta{P_0}=Nv$, that is, the final momentum of the fluid is such that the relative velocity between the stirrer and the fluid is minimised.   Correspondingly, the ensemble momentum (dashed line) as averaged over 248 trajectories approaches $\Delta{P}\approx\Delta{P_0}$.

Figures~\ref{fig:trajectories_long}(d--f) plot the momentum change for a random sample of thirty trajectories from the ensemble for the same three different stirring velocities to give an indication of the behaviour of individual trajectories.

Note in our simulations, the system does not always reach exactly $\Delta{P_0}$ as the initial stirrer speeds are not generally integer multiples of the momentum quantum $2{\pi}N/L$.

\subsection{Comparison with mean-field theory}
In this section we make a direct comparison between the TWA simulations and mean-field theory. We calculate the total momentum change $\Delta{P}$ as a function of initial stirrer speed for a total simulation time of $t_f=40L/v$, corresponding to 40 periods of stirring. We vary the stirring speed across the critical velocity $v~\in~[0.8-1.2]v_c$

For stirrer velocities with $v>v_c$ it is not possible to initialise the system with the stirrer already immersed in the fluid as no mean-field solution exists.  For this reason, in this section we begin the simulations without the obstacle present, and increase the amplitude of the potential $V_0$ smoothly as a function of time from zero to $0.3\mu$ as
\begin{equation}
    V_0(t) =  0.3\mu \sin^2{(\pi{t}/2\tau_r)}.
\end{equation}
The ramping period $\tau_r$ is of order $10^{-3}$ to $10^{-2}$ times the simulation interval. It results in transient and small amplitude dispersive waves which lead to a small change in the momentum of the fluid, but do not otherwise affect simulation outcomes~\footnote{The transient momentum change described here has no significant effect on the results presented, and could be further minimised a longer ramp period. We have independently verified using a time-dependent Hartree-Fock-Bogoliubov simulations that such a process does not seed dynamic instabilities in the Bogoliubov modes.}.  The regime $v \gg v_c$ is outside the scope of this work, but we note previous work that indicates that at sufficiently large velocities the obstacle no longer creates solitons~\cite{Radouani,Pavloff,Katsimiga2018}.

\begin{figure}
\includegraphics[width=0.90\textwidth]{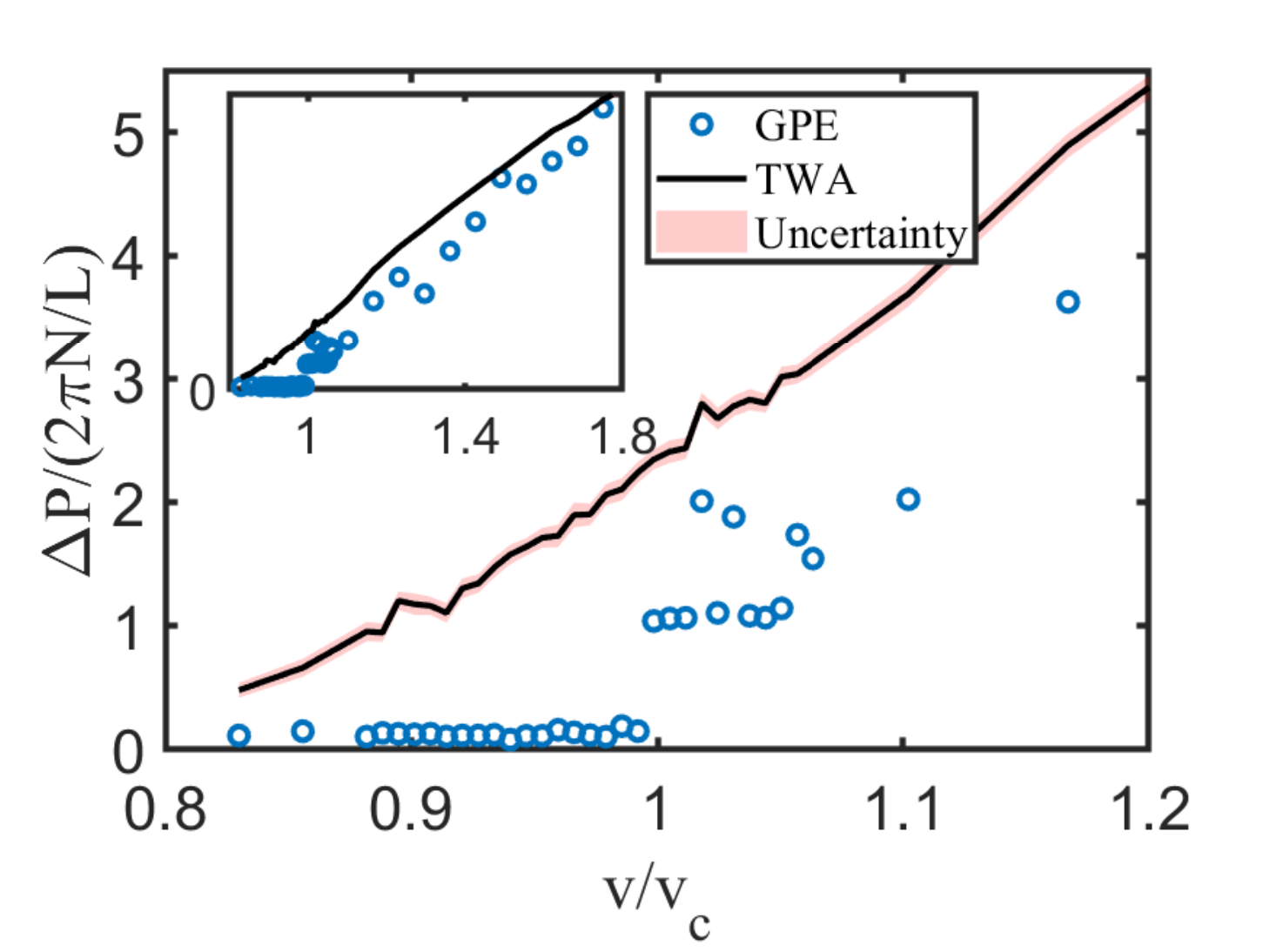}
\caption{Comparison of the momentum change for GPE (blue open circles) and TWA (black line) simulations as a function of stirrer velocity for a simulation time of $t_f=40L/v$.
Other simulation parameters are $g=0.1$, $N=1024$, $L=128$ using $128$ grid points and $V_0=0.3\mu$. 
The inset shows the same results for a wider range of stirrer velocities $v \in [0.6-1.8]v_c$, where in these simulations $v_c=0.38555$. The red shaded region indicates the 95\% confidence interval uncertainty for the TWA result for an ensemble of 96 trajectories.   The fluctuations in the GPE result as a function of $v$, and in the TWA simulations beyond the statistical uncertainty, is due to finite simulation time effects, such as when solitons and sound waves deterministically collide with the obstacle or each other, resulting in small, transient momentum changes. They do not qualitatively affect the main observation of a sharp step at $v/v_c=1$ in the mean-field results but no such step change occurs for the TWA simulations.
\label{fig:above_vc}}
\end{figure}

The momentum increase for these simulations is plotted in Fig.~\ref{fig:above_vc}.  In the mean-field simulations, where the initial fluid speed is below the critical velocity ($v < v_c$), there is no increase in the momentum of the fluid beyond the initial transients~\cite{Hakim,Pavloff,Feng}. At $v=v_c$ there is a step-like increase in $\Delta{P}$ corresponding to the onset of soliton formation. 

In contrast, in the TWA simulations $\Delta{P}$ increases continuously across $v = v_c$. Within the discretisation of the simulated stirrer speeds, $\Delta{P}$ varies smoothly with no suggestion of a threshold. 
Below $v_c$, $\Delta{P}$ falls toward zero as $v/v_c\rightarrow 0$. We can understand this qualitatively as follows: for $v<v_c$, transitions between the two quantised flow states of the fluid are classically forbidden and are only possible through quantum tunnelling. As $v\rightarrow{v_c}$, even the smallest fluctuations can cause an irreversible transition to a superflow state with a lower energy in the rotating frame. 

Above $v_c$, soliton formation is energetically allowed in mean-field theory and occurs with certainty in both TWA and GPE simulations at an average rate that increases with $v$, consistent with previous work~\cite{Hakim,Pavloff,Astrakharchik}. The TWA  result has a slightly larger momentum change compared to GPE simulations for the same stirrer velocity. This is most noticeable for $v\gtrsim{v_c}$. For $v\gg{v_c}$ the momentum change appears to be dominated by mean-field physics and effects due to quantum fluctuations are small in comparison (see inset to Fig.~\ref{fig:above_vc}).

These simulations were performed in a parameter regime where interactions are relatively strong ($g=0.1)$ such that excitations below $v_c$ occur more readily. We have also performed simulations for smaller values of $g$ with qualitatively similar results, although $\Delta P$ falls off more rapidly below $v_c$.

\subsection{Long time behaviour}

\begin{figure}
\includegraphics[width=1.0\textwidth]{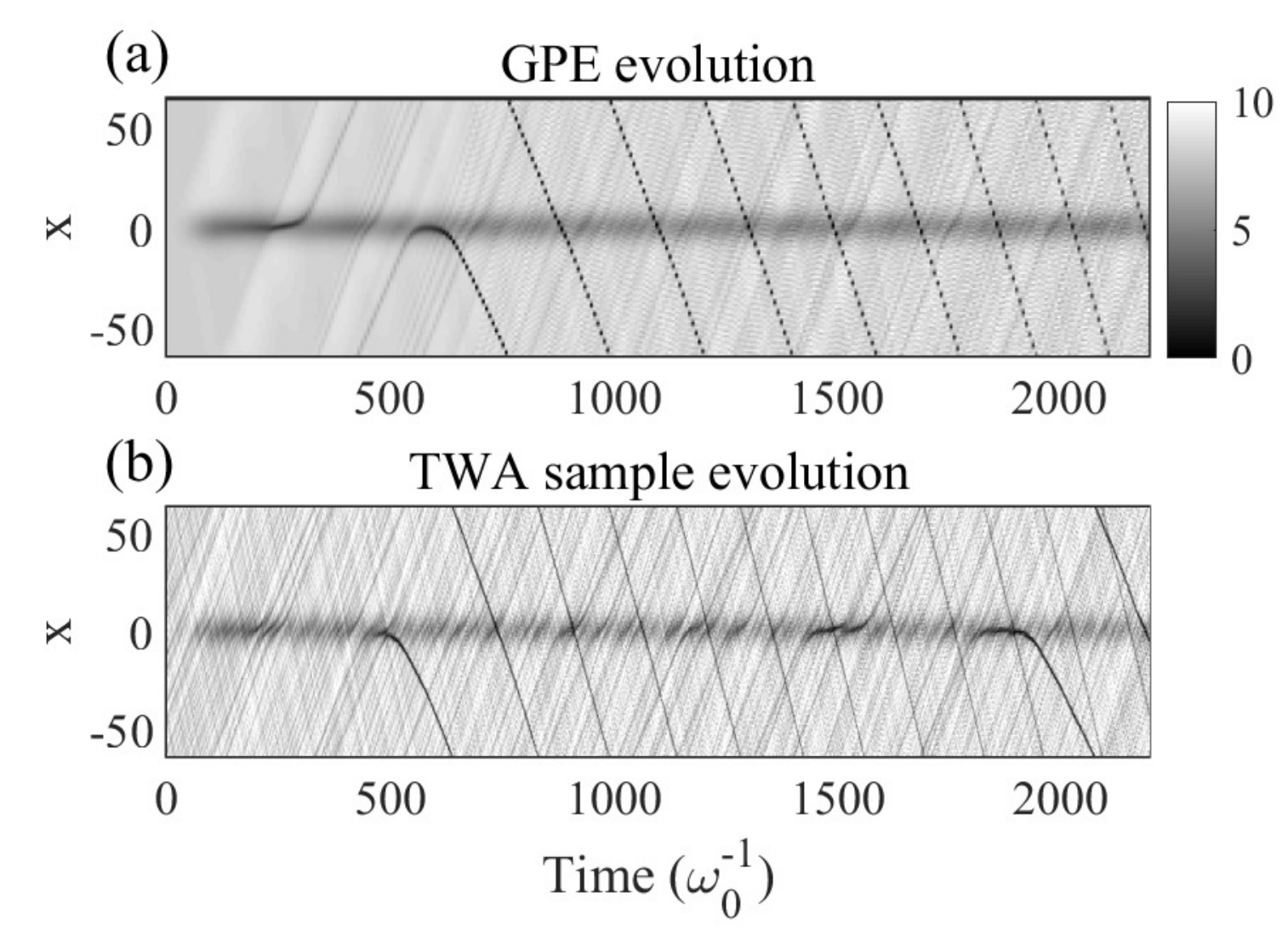}
\caption{Example comparison of GPE and TWA densities $|\psi(x)|^2$ for stirring with $v>v_c$. (a) GPE simulation for which soliton formation ceases after the first soliton, which itself persists for some time. (b) TWA simulation, where the first soliton decays over time (becomes less dark) and a second soliton forms toward the end of the simulation interval. Simulation parameters are $g=0.1$, $N=1024$, $L=128$ using $128$ grid points and $V_0=0.3\mu$. Here, $v=1.0155v_c$, $\tau_s=235$.
}
\label{fig:finite_size}
\end{figure}

We observe distinct differences in the long-time behaviour of the GPE and TWA simulations. An illustration of this is shown in Fig.~\ref{fig:finite_size}.  Figure~\ref{fig:finite_size}(a) shows the density as a function of time for a GPE simulation for $v>v_c$. Figure~\ref{fig:finite_size}(b) shows a typical trajectory for a TWA simulation for the same parameters --- we see that only a single soliton forms in the GPE simulation, while a second soliton forms in the TWA simulation. 

For GPE simulations with a stirring speed above $v_c$, we find that after a sufficient number of solitons form the relative speed between the fluid and the stirrer drops below $v_c$. At this stage further soliton formation appears to be suppressed. Quantitatively, the rate of change of momentum peaks and falls off after characteristic time $\tau_s \approx L/(c-v)$, which corresponds to the time it takes for the obstacle to catch up with forward propagating waves from when it was first introduced. 

In contrast, this suppression of excitations does not occur in the TWA simulations, as illustrated in the sample trajectory shown in Fig.~\ref{fig:finite_size}(b).  Instead, the formation of the first soliton appears to trigger a cascade of additional solitons until the relative speed between the stirrer and the fluid is close to zero (note Ref.~\cite{Lychkovskiy2015} suggests the long time velocity can be finite). The time intervals between subsequent soliton formation events are stochastic, but are typically significantly shorter than the average time to formation of the first soliton --- see e.g. Fig.~\ref{fig:trajectories_long}.

\begin{figure}
\includegraphics[width=0.92\textwidth]{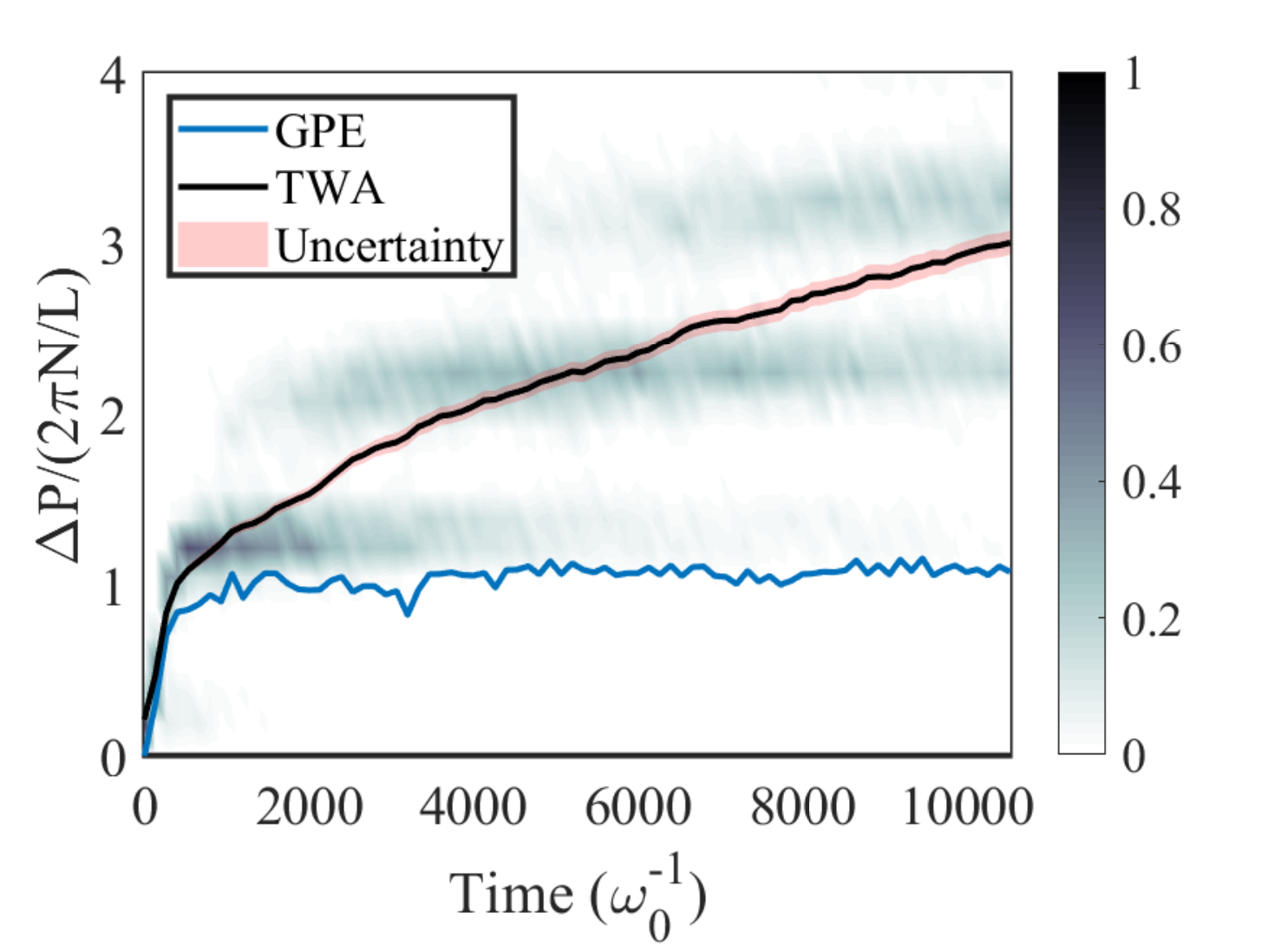}
\caption{Comparison of the change in momentum for a GPE simulation (blue line) and the TWA ensemble (black line) averaged over 96 trajectories, the red shaded regions represent the 95\% confidence interval. Simulation parameters are as in Fig.~\ref{fig:finite_size}. The fluctuations of the momentum as a function of time for the GPE result is due to the interaction of excitations with the obstacle. The background greyscale density plot shows a histogram of the fraction of TWA trajectories with each momentum (35 bins over $\Delta{P}/(2\pi{N}/L)\in[0,5]$).
\label{fig:finite_size_long}}
\end{figure}

Figure~\ref{fig:finite_size_long} compares the increase in the total momentum for the GPE and the TWA simulations for the same conditions as for Fig.~\ref{fig:finite_size}.  It is clear that the momentum reaches a plateau for the GPE corresponding to a single soliton, but the TWA simulation continues to form solitons and the momentum continues to increase. The physical mechanism for this difference in behaviour is unclear.  We note that the solitons in the TWA simulations decay over time (become less dark), which does not seem to occur in GPE simulations. We leave this observation as a topic for future study.

\subsection{Tunnelling rate between quantised flow states}\label{sec:decay_rate}

Calculating the decay rate of a persistent current has been a focus of previous work, and it is a key observable in experiments. A characteristic result of such calculations~\cite{Buchler,Kagan,Cherny,Cherny_Review,Danshita} is that for the highest order transition, from a winding number of $q$ to $q-1$, the transition rate (and superfluid decay rate) $\Gamma$ has a characteristic power-law dependency on the obstacle velocity $v$ of
\begin{equation}
\Gamma \propto (v/v_c)^{\alpha K - \beta},
\label{eq:scaling}
\end{equation}
where 
\begin{equation}
K = \frac{\pi}{\sqrt{4\pi a_s \hbar L/N}},
\label{eq:tomonaga_luttinger}
\end{equation}
is the Tomonaga-Luttinger parameter which characterises the strength of interactions in the one-dimensional system~\cite{Buchler,Cherny_Review}. It is commonly expressed in terms of the Lieb-Liniger gas parameter $\gamma = 4\pi a_s \hbar L/N$ ($a_s, L$ and $N$ are defined as before) such that $K = \pi/\sqrt{\gamma}$. 

The calculations agree that $\alpha = 2$, and $\beta$ is an integer of order unity that varies depending on the system geometry and approximation regime used~\cite{Buchler,Kagan,Cherny,Danshita}. For ring systems at ${T=0}$ perturbed by a delta-like impurity, the effective Hamiltonian method results in $\alpha=2$ and $\beta=1$~\cite{Kagan,Cherny_Review}. For infinite systems at $T=0$ perturbed by a delta-like impurity, the instanton method gives similar results~\cite{Buchler}. For lattice potentials, an approach using the Bose-Hubbard model at $T=0$ indicates $\alpha=2$, $\beta=2$~\cite{Danshita}. A summary of these results is provided in Table~\ref{tab:summary_scaling}.

\begin{table*}
\begin{tabular}{|l|l|l|l|l|}
\hline
Study & Approach & Temperature & Geometry & Result \\
  \hline
Kagan 2000 \cite{Kagan} & Effective Hamiltonian & $T=0$ &  Ring with delta impurity & $\Gamma \propto v^{2K-1}$ \\
Cherny 2011 \cite{Cherny} & Effective Hamiltonian & $T=0$ & Ring with delta impurity, $v \ll v_c$ & $\Gamma \propto v^{2K-1}$ \\
B\"{u}chler 2001 \cite{Buchler} & Instanton & Finite $T$& Infinite system with delta impurity & $\Gamma \propto v T^{2K-1}$ \\
		&  & $T=0$ & Infinite system with delta impurity & $\Gamma \propto v^{2K-1}$ \\
Danshita 2012 \cite{Danshita} & Bose-Hubbard model & $T=0$ & Lattice potential & $\Gamma/L \propto v^{2K-2}$ \\
	    &  & Finite $T$ & Lattice potential & $\Gamma/L \propto v T^{2K-3}$ \\
\hline
\end{tabular}
 \caption{Calculations of the superfluid flow decay rate $\Gamma$ by various methods. 
 }
 \label{tab:summary_scaling}
\end{table*}

We note that Eq.~(\ref{eq:scaling}) is a calculation of the transition rate for the first decay from $q$ to $q-1$ only. In the $K\gg~1$ regime, this is the dominant transition as higher quanta decays (e.g. $q\rightarrow q - n$ for $n>1$) are suppressed~\cite{Kagan,Cherny_Review}.

\begin{figure}
\includegraphics[width=1.0\textwidth]{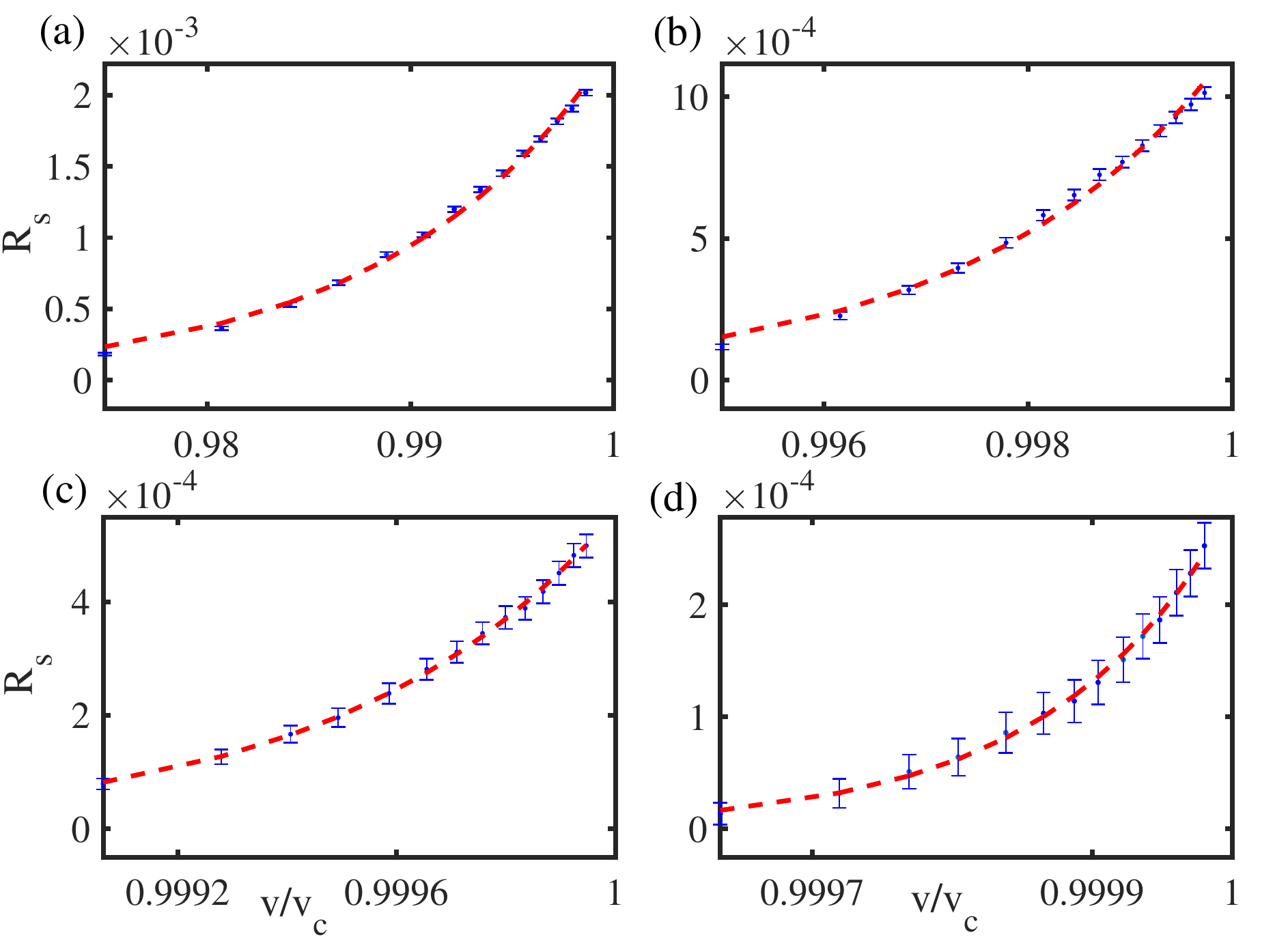}
\caption{Soliton formation rate $R_s$ as function of velocity for (a) $K=112$ (b) $K=375$, (c) $K=1332$ and (d) $K=4683$. Simulation data is shown as blue markers, a fit to a power law function is shown as dashed red lines. System parameters ranged from $g\in[2.5\times10^{-2},6\times10^{-4}]$ and $N\in[16384,682667]$, with $L=512$ using $512$ grid points and $V_0=0.3\mu$. Each data point was obtained from sampling 3000 trajectories. Error bars representing 95\% confidence intervals are shown in all figures: for (a--c) they are small and of similar size to the plot markers.
\label{fig:momentum_scaling}}
\end{figure}

We have tested these analytically derived scalings numerically  through simulations for a broad range of $K \sim 112$--$5100$ ($\gamma \sim 10^{-5}$--$10^{-7}$) where the validity conditions for the truncated Wigner approach are satisfied, and which correspond to  parameters that are potentially experimentally accessible. We adjust $K$ in the numerics by systematically decreasing the interaction strength $g$ and increasing the atom number $N$ while keeping $gN/L$ constant. This keeps the mean-field dynamics unchanged, while reducing the relative amplitude of the quantum fluctuations in the TWA simulations. For each value of $K$, we perform simulations for an appropriate range of sub-critical obstacle velocities, typically $v=0.8-0.9999 v_c$ and determine the decay rate for each set of parameters using an ensemble of 3000  trajectories. Simulation parameters were $L=512$, $\sigma=3\xi$, $V_0=0.3\mu$ and $M=256$, with $gN/L=0.8$ and $v_c=0.44098c=0.39442$. We restrict our attention to short times when the decay should be dominated by the first transition $q\rightarrow~q-1$. As $K\gg 1$ in our simulations and $\beta$ is of order unity, we compare our results to the approximate scaling relation $\Gamma \propto \left(v/v_c\right)^{\alpha K}$,

Two different methods are used to estimate the decay rate $\Gamma$. In the first, we track the rate of soliton formation per unit time in individual trajectories and use these to calculate the ensemble average soliton formation rate $R_s$ for a range of velocities $v<v_c$. Given the correlation between soliton formation and the momentum change, $R_s$ is a suitable proxy for the transition rate $\Gamma$. We fit this to the power-law relationship $R_s \propto \left(v/v_c\right)^{\alpha K}$. Sample data and fits are shown in Fig.~\ref{fig:momentum_scaling} for different parameter regimes.  We find that the numerical results are well-fit by this function.

\begin{figure}
\includegraphics[width=0.75\textwidth]{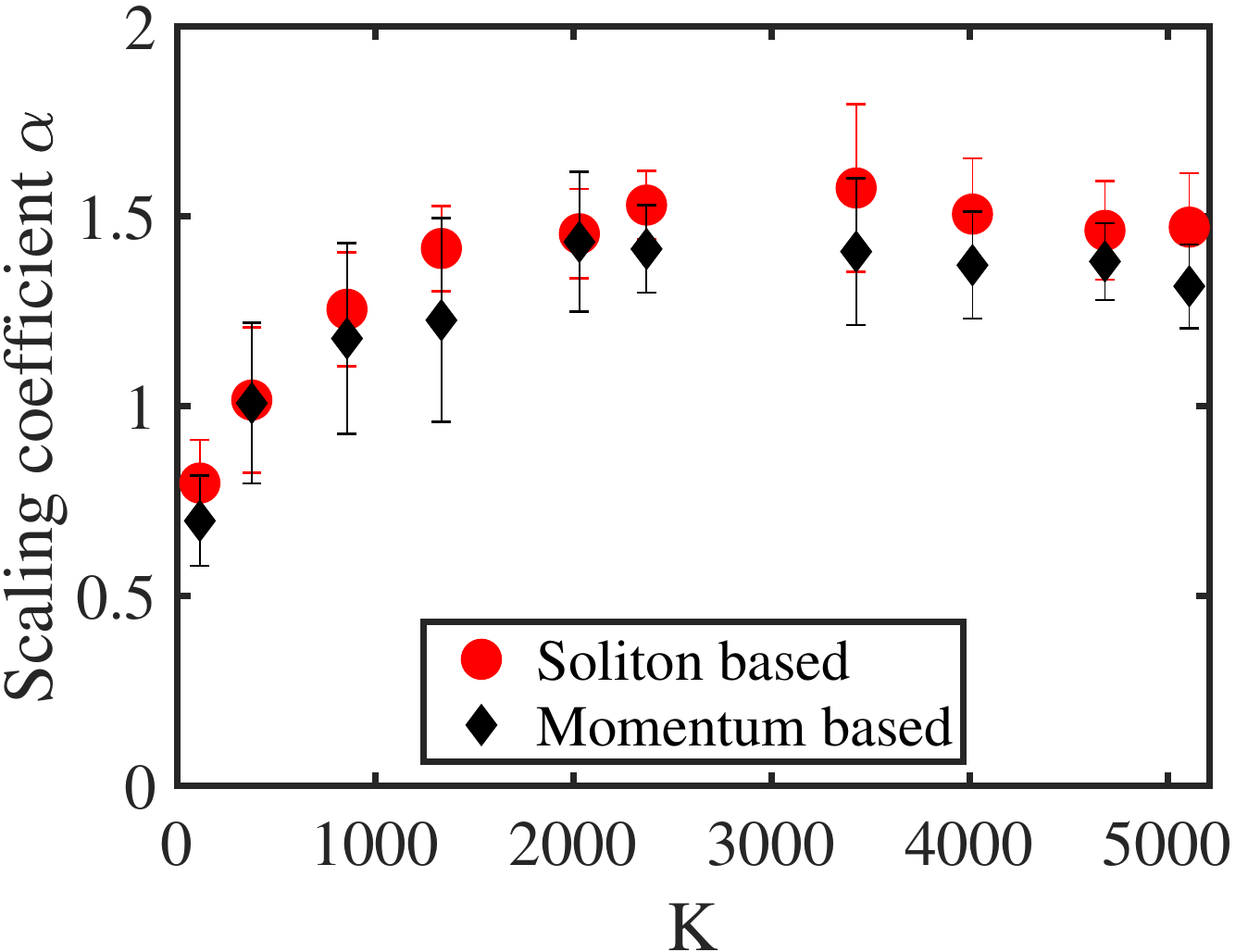}
\caption{Scaling exponent $\alpha$ as a function of the Tomonaga-Luttinger parameter $K$ found by measuring the rate of soliton formation (red circles) and the rate of momentum decay (black diamonds). The range of parameter values used in the TWA simulations are as described in the caption to Fig. \ref{fig:momentum_scaling}. Error bars represent 95\% confidence intervals.\label{fig:scaling_coefficient}} 
\end{figure}

The exponent $\alpha{K}$ is estimated during the fitting process using a standard weighted regression algorithm in the MATLAB software suite, and  is plotted as red circles in Fig.~\ref{fig:scaling_coefficient}. We find that $\alpha\approx1.5\pm0.1$ in the weakly interacting regime ($K=5109$) and drops to  $\alpha\approx0.8\pm0.1$ for stronger interactions ($K=112$). These results for $\alpha$ are statistically smaller than the theoretical predictions of $\alpha=2$ as reported in Refs.~\cite{Buchler} which is valid for $K>1$ and in Refs.~\cite{Kagan,Cherny_Review,Danshita}, valid in the weakly interacting regime $K\gg{1}$. 

The expression $\Gamma \propto \left(v/v_c\right)^{\alpha K-\beta}$ assumes the overall decay of the system is dominated by the single quanta decay of its highest occupied state. This assumption is satisfied for our simulations with $K>10^3$ where fewer than $1\%$ of trajectories are observed to have formed more than a single soliton within the simulation interval. However, for smaller $K$ this assumption is less reliable as decay is influenced by higher rates of subsequent soliton formation, including the cascade effect described earlier. For simulations at $K=374$ and $K=112$, up to $10\%$ and $77\%$ of trajectories recorded the formation of more than a single soliton during the simulation interval, and it seems likely that the smaller values of $\alpha$ we find at lower values of $K$ are due to this effect.  

We have also performed a second estimate of the scaling of the transition rate with $v$  by calculating the rate of change of the ensemble momentum over a fixed interval, excluding a short initial period of small and transient momentum changes. For the relatively short simulation intervals used in these sets of simulations, the behaviour is approximately linear, of form $\Delta{P}(t)\propto \Gamma{t}$. We estimate $\Gamma$ from the gradient for a range of velocities $v<v_c$ and fit to the expected power relationship as before to estimate $\alpha$. The linear fits have coefficient of determination ($R$-squared) values between 0.97 (at lower velocities) and 0.99 (at higher velocities), indicative of a good fit.

Again, the exponent $\alpha{K}$ is estimated from the fits and the results are shown as black diamonds in Fig.~\ref{fig:scaling_coefficient}. We find $\alpha\approx0.7\pm0.1$ for small $K$ and $\alpha\approx 1.3\pm0.1$ at large $K$, similar to the soliton-based measure.

The difference between the predicted scaling law and our findings is currently unexplained.  Our estimate of $\alpha$ is based on measurements over a range where $\Gamma$ varied by a factor of ~45. A more reliable practice for fitting data to power law expressions would be to consider a wider range. However, going further is beyond the available computational resources as it requires a factor of ten increase in the number of simulations, or a commensurate increase in the integration time of each simulation.  Another potential factor in the difference is the type of obstacle used for stirring.  In the theoretical predictions this was a microscopic delta-function potential, or a lattice potential, whereas here we have used a macroscopic Gaussian potential several times larger than the healing length.  
  Notwithstanding the limitations of our simulations, the observed scaling is similar to the theoretically predicted power law form.  Our simulations have demonstrated that in the TWA approach, the macroscopic tunnelling of quantised flow states is directly linked to the stochastic formation of solitons in the microscopic dynamics.  

\subsection{Discussion}
Our results suggest that observing the decay of a quantised current due to quantum fluctuations will be challenging in experiments  in the weakly interacting regime ($K\gg1$), as the rate decreases rapidly for $v<v_c$ and may only be experimentally observable very close to $v_c$. For experiments in the more strongly interacting regimes ($K\gtrsim{1}$), our results show that soliton excitations may be observable even at velocities significantly below the mean-field critical velocity. To observe the spontaneous formation of dark solitons below the critical velocity,  experiments would need to be able to differentiate momentum changes of individual quanta, and control the initial stirring obstacle velocity at a level of precision of order $10^{-2}$ to $10^{-3}$ times $v_c$. While challenging, this may be achievable in modern experiments~\cite{Tanzi}. 

Previous work using the TWA suggested quantum and thermal fluctuations could lead to a lower critical velocity compared to mean-field predictions~\cite{Mathey}. Our findings provide further insight --- while solitons may form at arbitrarily low velocities, the rapid fall off in the rate of excitations as $v\rightarrow0$, could manifest in experiments and numerical simulations with limited spatial resolution as an \emph{apparent} reduction in the effective critical velocity. This would be especially pronounced in experiments operating in regimes of weak interactions.

The change in quantised flow in 1D systems occurs due to the formation of solitons. Therefore, the statistics of soliton formation could provide an alternative method to track and measure the change of quantised flow. Solitons, and their higher dimensional analogues, are well-defined excitations that can be visible with high-resolution experimental imaging~\cite{Engels}. 

\section{Conclusions}
We have simulated the stirring of a zero temperature one-dimensional Bose gas with periodic boundary conditions by an obstacle above and below the mean-field critical velocity using the beyond-mean-field Truncated Wigner approximation. We have shown that the quantum fluctuations in the TWA result in probabilistic formation of grey solitons in the fluid below the critical velocity.

We have quantified changes in the superfluid flow both by tracking soliton formation, as well as calculating the change in the system momentum, and have observed a non-zero probability for soliton formation, and hence superfluid decay, at all obstacle speeds. We find that this probability decreases smoothly as the stirrer speed falls below $v_c$.  These results are distinguished from the mean-field case, where no decay is observed for $v<v_c$.

The average time to the formation of the first soliton depends strongly on the stirrer speed and the strength of interactions $g$. Within the TWA,  the formation of the first soliton triggers a soliton cascade which continues until the relative speed between the stirrer and the fluid is minimised. Such a soliton cascade is not observed in GPE simulations.

Finally, for $v<v_c$ we found the rate of formation of solitons, and the rate of change of momentum of the system, to scale as a power law of the fractional velocity $v/v_c$, consistent with the theoretically prediction $\Gamma \propto \left(v/v_c\right)^{\alpha K}$, where in our simulations we find $\alpha \approx (0.8-1.5)\pm0.1$. In the weakly interacting regime, we find $\alpha=1.5\pm0.1$ which does not agree with the results of $\alpha=2$~\cite{Kagan,Cherny_Review}.  A potential reason for this difference is the nature of the obstacle used in the stirring simulations. 

\begin{acknowledgement}
The authors gratefully acknowledge many insightful discussions with Tod Wright.  CF would like to acknowledge computational support from Graham Dennis. This research was supported by Australian Research Council Discovery Projects Nos. DP1094025 and DP110101047, and the Australian Research Council Centre of Excellence in Future Low-Energy Electronics Technologies (project number CE170100039) and funded by the Australian Government.
\end{acknowledgement}

\section*{Author contribution statement}
Both authors contributed to the development of the ideas and methods of analysis for the project.  C.F. performed all the simulations and data analysis, and wrote the first draft of the manuscript.  Both authors discussed and interpreted the results, and contributed to the writing of the final version.

\bibliographystyle{epj}

\bibliography{thesis_bib}
\end{document}